\documentclass[conference]{IEEEtran}
\IEEEoverridecommandlockouts

\usepackage{cite}
\usepackage{amsmath,amssymb,amsfonts}
\usepackage{algorithmic}
\usepackage{graphicx}
\usepackage{textcomp}
\usepackage{xcolor}
\usepackage{booktabs}
\usepackage{float}
\usepackage{tabularx}
\usepackage{array}
\usepackage{stfloats} 
\renewcommand\IEEEkeywordsname{Keywords}
\begin{document}

\title{A Novel 8T SRAM-Based In-Memory Computing Architecture for MAC-Derived Logical Functions}

\author{
\IEEEauthorblockN{Amogh K M}
\IEEEauthorblockA{PES University\\
amoghkaipa@gmail.com}
\and
\IEEEauthorblockN{Sunita M S}
\IEEEauthorblockA{PES University\\
sunitha@pes.edu}
}

\maketitle

\begin{abstract}
This paper presents an in-memory computing (IMC) architecture developed on an 8×8 array of 8T SRAM cells. This architecture enables both multi-bit parallel Multiply-Accumulate (MAC) operations and standard memory processing through charge-sharing on dedicated read bit-lines. By leveraging the maturity of SRAM technology, this work introduces an 8T SRAM-based IMC architecture that decouples read and write paths, thereby overcoming the reliability limitations of prior 6T SRAM designs. A novel analog-to-digital decoding scheme converts the MAC voltage output into digital counts, which are subsequently interpreted to realize fundamental logic functions including AND/NAND, NOR/OR, XOR/XNOR, and 1-bit addition within the same array. Simulated in a 90 nm CMOS process at 1.8 V supply voltage, the proposed design achieves 8-bit MAC and logical operations at a frequency of 142.85 MHz, with a latency of 0.7 ns and energy consumption of 56.56 fJ/bit per MAC operation and throughput of 15.8 M operations/s.
\end{abstract}


\section{Introduction}
Movement of data between the processor and memory creates a significant bottleneck in Von-Neumann architectures, primarily due to the high energy consumption introduced during data transfers. The energy consumption during transfer of data is greater than a hundred times the energy consumed during an arithmetic operation \cite{b1}. As a result, the traditional Von-Neumann architecture is heavily limited by overhead associated with data transfers. To address this issue, techniques such as in-memory computing \cite{b2} have gained popularity. IMC mitigates the bottleneck by enabling computation directly within the memory array, thus eliminating the additional energy associated with shuttling data between memory and processor. SRAM, being a commercially mature technology already integrated into computing systems of all sizes and types \cite{b3,b4}, makes the SRAM-based IMC approach capable of significantly reshaping the computing industry.

Embedding computation such as MAC within the memory element removes the long-standing divide between processor and memory. This offers a game-changing advantage for AI and deep learning tasks, where algorithms require extensive memory access \cite{b5}. IMC implementations using standard 6T SRAM cells provide compatibility with advanced CMOS processes and high storage density. In \cite{b6}, multi-bit multiplication within a 6T SRAM array is achieved by leveraging the controlled discharge of the bit-lines. When multiple word lines in a 6T SRAM are activated at the same time, the read noise margin drops and unintended short-circuit paths can occur, which may randomly flip the stored data \cite{b4}. In addition, because 6T cells share the same path for reading and writing, activating another word line after the bit line has discharged can unintentionally overwrite the stored value. These effects reduce the reliability of 6T SRAM for multi-operand logic and MAC operations.

Using an 8T bitcell offers a practical solution to the drawbacks of 6T SRAM \cite{b8}. By decoupling the read and write paths, 8T SRAM enables independent tuning of these operations, which enhances stability margins, supports lower Vdd operation, and ultimately reduces power consumption. This work introduces an 8T SRAM-based IMC architecture that derives AND/NAND, OR/NOR, XOR/XNOR, and 1-bit addition directly from the MAC discharge behavior, with no additional logic circuitry. An 8×8 array was constructed using 8T SRAM cells. Specifically, the proposed architecture can implement M parallel N-bit MAC operations. Logical operations such as AND/NAND, NOR/OR, XOR/XNOR, and 1-bit addition operations have been derived using a novel decoding scheme.

The rest of the paper is organized as follows: Section II deals with implementation and design of the 8T SRAM cell, the 8×8 array architecture, and the peripheral circuitry. Section III explains MAC and logical operations derived from MAC. Section IV presents the results, and Section V concludes the paper.

\section{8T SRAM CELL AND ARRAY ARCHITECTURE}

\subsection{Key Idea}
The proposed architecture employs an 8×8 array of 8T SRAM cells that supports both conventional memory operations and in-memory computing capabilities. While standard read and write operations are performed through traditional peripheral circuitry, the MAC operations leverage the charge-sharing characteristics of the dedicated read bit-lines (RBL). When multiple word lines are activated simultaneously, RBL voltage drops proportionally to the number of active cells, creating an analog representation of the MAC result. A custom MAC decoder circuit then digitizes this voltage level to produce the corresponding MAC count. Building upon this foundation, this work demonstrates how logic operations AND/NAND, OR/NOR, and XOR/XNOR can be directly inferred from the MAC results using novel interpretation techniques. This effectively transforms a single MAC operation into a versatile logic processing unit capable of performing multiple computational tasks within the same memory array.

\begin{figure}[H]
    \centering
    \includegraphics[width=\columnwidth]{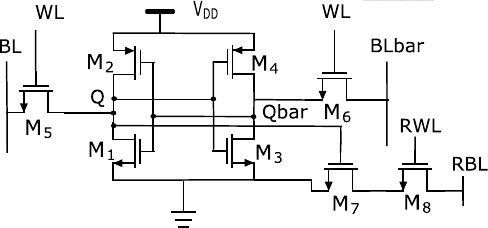} % <-- replace with your file name
    \caption{8T SRAM cell architecture.}
    \label{fig:cell}
\end{figure}

\begin{figure*}[!b]
    \centering
    \includegraphics[width=\textwidth, height=0.45\textheight, keepaspectratio]{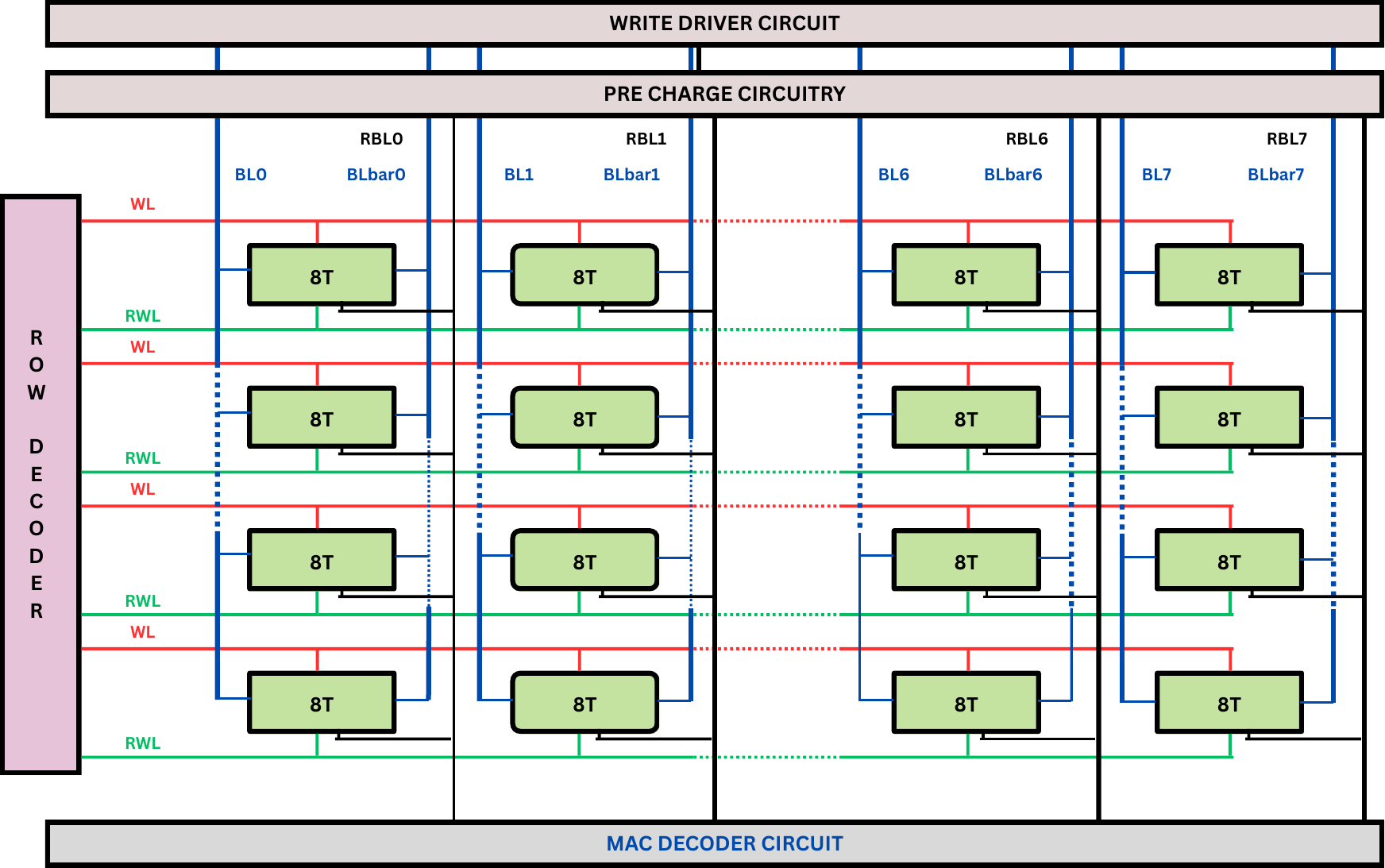} 
    \caption{8×8 IMC array.}
    \label{fig:array}
\end{figure*}

\subsection{8T SRAM Cell Architecture}
The 8T SRAM cell, shown in Fig.~\ref{fig:cell}
, is the primary building block of the 8×8 memory array. It consists of a conventional 6T SRAM cell and two back-to-back NMOS transistors M7 and M8 attached to node Q. M7 and M8 are called the read buffer and read access transistors, respectively. The SRAM cell is written using BL and BLbar lines, which are driven by the write driver circuit. To ensure the data in the cell is not flipped, the widths of M1 and M3 transistors are twice the widths of the other transistors. The Read Word Line (RWL) and RBL signals control the access to node Q through M7 and M8.

On connecting node Q to the gate of M7 in the 8T SRAM cell, the data ‘1’ stored in the cell leads to a maximum drop of RBL on activating RWL. This is because there is a discharge path created from RBL to ground via M7 and M8, causing the RBL voltage to drop. On the other hand, when the cell stores ‘0’, there is no discharge path from RBL to ground since M7 is disabled, and the RBL voltage remains at the pre-charge value. The advantage of connecting node Q to the gate of M7 is that the MAC operation results in a drop in RBL voltage level that is proportional to the MAC count as explained in Section III.

\begin{figure}[!t]
    \centering
    \includegraphics[width=0.8\columnwidth]{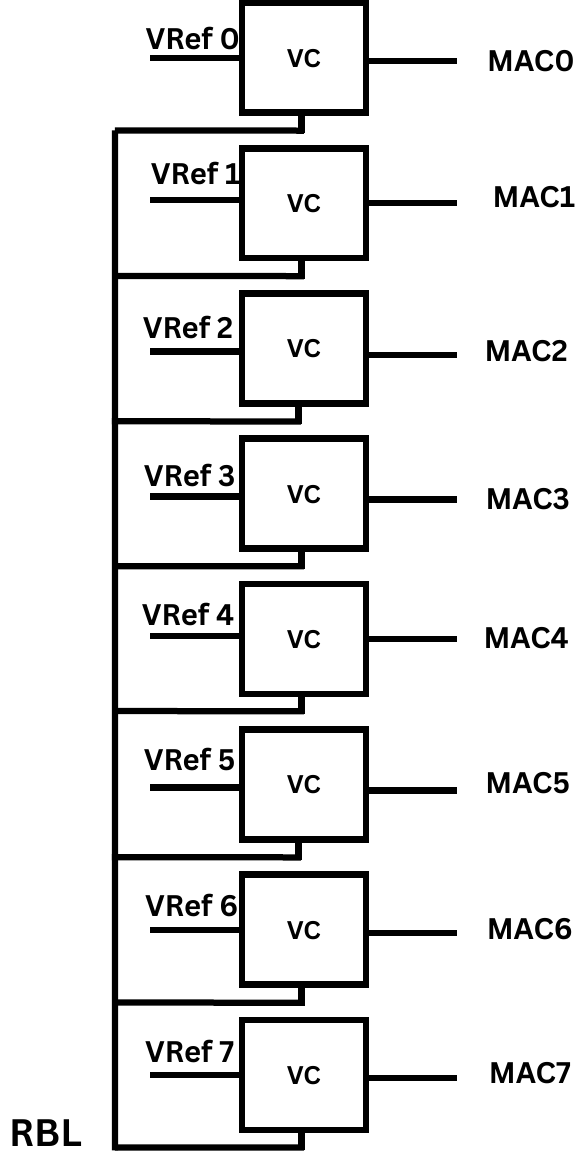} % rename as needed
    \caption{MAC decoder circuit}
    \label{fig:mac_decoder}
\end{figure}

\subsection{8×8 IMC Array}
A block representation of an 8×8 array is shown in Fig. 2. Each row has shared WL and RWL signals. Each column has shared BL, BLbar, and RBL signals. The MAC operation is achieved using the charge-sharing capability of dedicated RBLs. This method ensures there is no read disturbance since there is a dedicated read bit-line, separate from the write bit-line.

The 8×8 array consists of the peripheral circuits, which include the write driver, pre-charge circuits, row/column decoders, and MAC decoder circuits. The write driver circuit connects to the cells through multiplexers. One pre-charge circuit is present per column, which pre-charges the RBL lines. The row and column decoders are used to select the cell to which the data has to be written or read. They consist of 3:8 decoders. The multiply and accumulate operations are performed on the stored operands by exploiting the charge-sharing capabilities of the read bit-line of each column.

\subsection{MAC Decoder Circuit}
The MAC decoder circuit is used to decode the RBL voltage values and interpret the result by digitizing the RBL voltage. It consists of eight voltage comparators, as shown in Fig. 3. The voltage threshold of each comparator is set according to the RBL voltage corresponding to each MAC result. Each column has one MAC decoder circuit.

The circuit of the voltage comparator (VC) used in the decoder unit is shown in Fig. 4. A seven-transistor voltage comparator using a two-stage differential amplifier has been used with default transistor sizes. The MAC decoder circuit is used for decoding the RBL voltage values and interpreting the MAC result by digitizing the RBL voltage.

\begin{figure}[!t]
    \centering
    \includegraphics[width=\columnwidth]{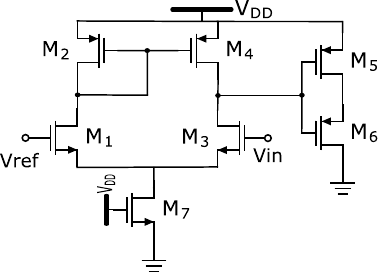} % rename file
    \caption{Voltage comparator circuit.}
    \label{fig:voltage_comparator}
\end{figure}

\section{MULTIPLY-ACCUMULATE AND LOGIC OPERATIONS}

\subsection{Multiply and Accumulate Operation}
In the proposed design, the Multiply and Accumulate (MAC) operation is performed directly inside the 8T SRAM array. Consider a MAC operation between operand A and operand B, both of which are 8 bits long. The bits of operand B are stored one after another in different rows of the same column during consecutive write cycles. After loading operand B, the Read Bit-Line (RBL) is pre-charged to 1.8 V. Then, the corresponding bits of operand A are applied to RWL on corresponding rows for a short duration.

The voltage on the RBL drops depending on how many bits of A and B match as logical high. Specifically, the more matched bits there are, the more SRAM cells discharge the RBL, causing a larger voltage drop. For example, a MAC count of 8 means all eight SRAM cells in that column contribute to discharging the RBL, while a count of 0 means none do, so the voltage remains at the pre-charged level. Intermediate counts from 1 to 7 produce proportional voltage drops, allowing the MAC result to be directly determined by measuring the RBL voltage. A table summarizing the relationship between MAC counts and RBL voltages is provided in Table~\ref{table:mac_vs_rbl}
. The MAC decoder circuit takes RBL as input and outputs an 8-bit signal that indicates the decoded MAC result.

\begin{table}[!t]
\centering
\caption{Relationship between RBL voltages and MAC count}
\label{table:mac_vs_rbl}
\label{tab:mac_rbl}
\begin{tabular}{ccccc}
\toprule
A Pattern & B Pattern & MAC & $V_{\text{RBL}}$ (V) & Decoded MAC Result \\
\midrule
xxxxxxxx  & 00000000  & 0 & 1.758 & 11111111 \\
1xxxxxxx  & 10000000  & 1 & 1.528 & 01111111 \\
11xxxxxx  & 11000000  & 2 & 1.308 & 00111111 \\
111xxxxx  & 11100000  & 3 & 1.096 & 00011111 \\
1111xxxx  & 11110000  & 4 & 0.895 & 00001111 \\
11111xxx  & 11111000  & 5 & 0.712 & 00000111 \\
111111xx  & 11111100  & 6 & 0.552 & 00000011 \\
1111111x  & 11111110  & 7 & 0.418 & 00000001 \\
11111111  & 11111111  & 8 & 0.310 & 00000000 \\
\bottomrule
\end{tabular}
\end{table}

In the proposed IMC architecture, the result of basic Logic operations can be directly inferred from the measured MAC count. Here, the MAC count simply represents the number of SRAM cells that actively discharge the RBL during evaluation. By choosing appropriate RWL activation patterns and input data configurations, different logic functions can be mapped to specific MAC count values as depicted in Table~\ref{table:logic_interpret}
.

\subsection{Interpreting AND/NAND from MAC Count}
Consider an intended one-bit AND/NAND operation (e.g., RWL pattern: 11, Data = 11). This results in a MAC count of 2. This indicates that both inputs are logic high, producing an AND result of 1. Any MAC count less than 2 signals that at least one operand bit is low, so the AND result is 0. The NAND output is simply the inverted AND result.

\subsection{Interpreting NOR/OR from MAC Count}
Consider an intended one-bit NOR/OR operation (e.g., RWL pattern: 11, Data = 00). This results in a MAC count of 0. This indicates that both inputs are logic low, resulting in a NOR output of 1. Any nonzero MAC count means at least one input is high, giving a NOR result of 0 and an OR result of 1.

\subsection{Interpreting XOR/XNOR from MAC Count}
Consider an intended one-bit XOR operation (RWL pattern: 11, Data = 10). This results in a MAC count of 1. This corresponds to exactly one operand being high (logical XOR = 1). If the MAC count is not 1, it means both inputs are the same (both ‘0’ or both ‘1’), giving an XOR result of 0.

\subsection{Implementation of 1-Bit Addition Operation}
Consider an intended one-bit addition operation between data stored in two cells of the same column. When the RWLs of the two rows are activated, the sum given by A XOR B can be determined using the “MAC 1” result, and the carry given by A·B can be determined using the “MAC 2” result.

\begin{table}[H]
\centering
\caption{Interpreting AND, NOR, XOR Logic Operations}
\label{table:logic_interpret}
\label{tab:logic}
\begin{tabular}{c c c c c c}
\toprule
\small Data & 
\small $V_{\text{RBL}}$ (V) & 
\small Decoded & 
\small AND/ & 
\small NOR & 
\small XOR/ \\
\small (bitwise) & & 
\small MAC & 
\small Carry & & 
\small Sum \\
\midrule
00 & 1.758 & 0 & 0 & 1 & 0 \\
01 & 1.528 & 1 & 0 & 0 & 1 \\
10 & 1.528 & 1 & 0 & 0 & 1 \\
11 & 1.308 & 2 & 1 & 0 & 0 \\
\bottomrule
\end{tabular}
\end{table}

\subsection{Scalability}

Although this work presents results on an 8×8 array, the underlying MAC to voltage behavior extends naturally to larger arrays. The RBL discharge follows a charge sharing trend that scales predictably with the number of activated cells. In the current design, adjacent MAC levels are separated by a range of 100 mV to 250 mV, which is well above the comparator’s input-referred noise. When the array size increases, this spacing reduces in proportion to the effective bit-line capacitance, but it remains distinguishable by appropriately selecting the reference thresholds in the MAC decoder. Therefore, scaling requires re-tuning the reference voltages, and does not impose a fundamental limitation on extending the array to larger dimensions.

\section{Results}
The proposed design is implemented in Cadence Virtuoso using a 90 nm CMOS process technology. Simulations confirm the design’s ability to perform 8-bit Multiply and Accumulate operations and interpret results of 8-bit AND(NAND), 8-bit NOR(OR), and 8-bit XOR(XNOR) operations. At a supply voltage of 1.8 V, the energy consumption for 8-bit operations is calculated to be 452.2 fJ, resulting in an energy efficiency of 56.56 fJ/bit.

\begin{figure}[H]
    \centering
    \includegraphics[width=\columnwidth]{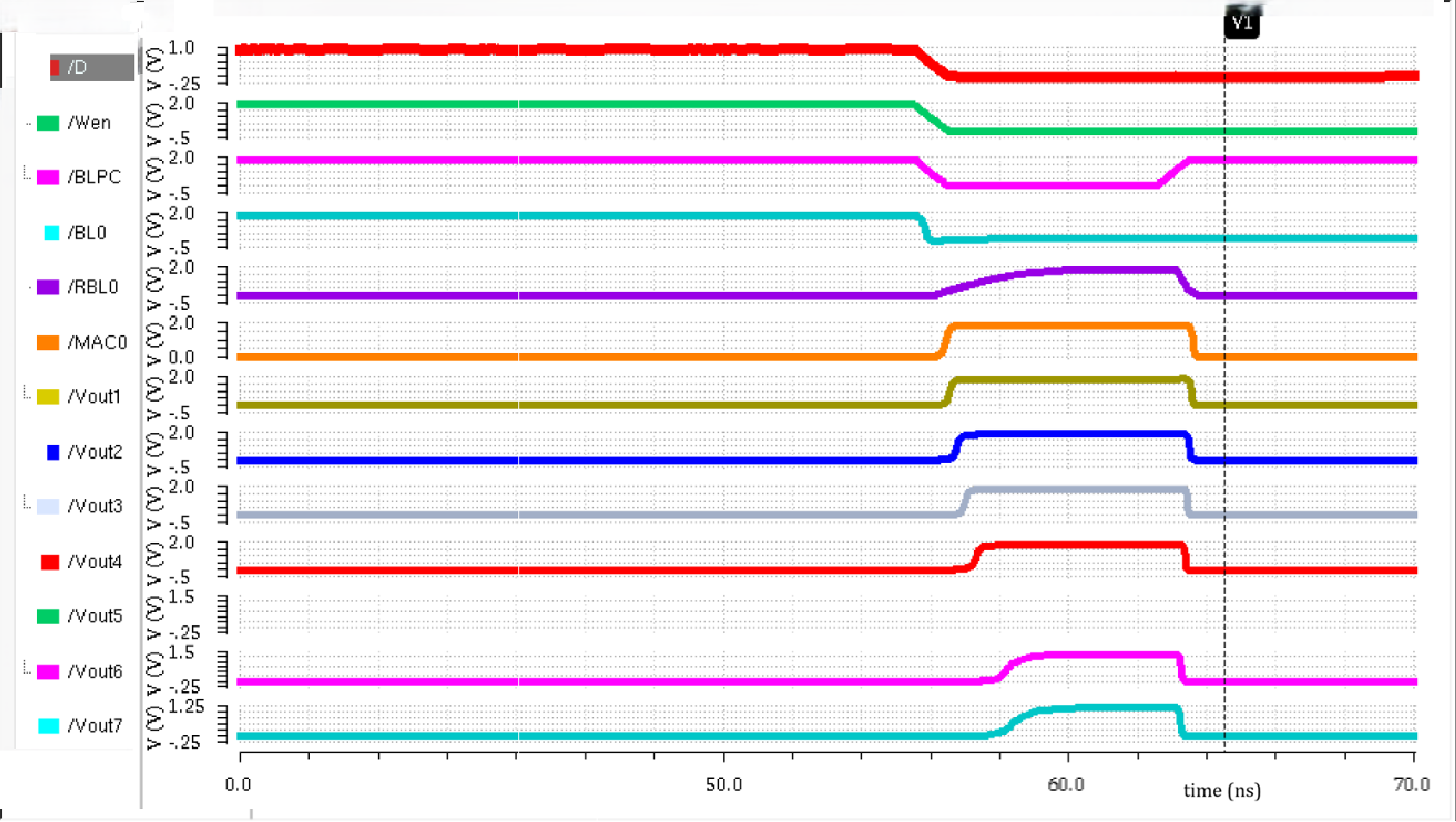} % rename to your file
    \caption{Output waveform for the 8-bit MAC operation}
    \label{fig:mac_waveforms}
\end{figure}

\subsection{MAC Simulation}
Fig.~\ref{fig:mac_waveforms} illustrates the MAC operation for two 8-bit operands, both equal to ‘11111111’. The bits of the second operand are sequentially written into eight rows of the selected column over eight write cycles. After the operand is loaded, the corresponding RBL is pre-charged to 1.8 V by asserting the BLPC signal. In the subsequent cycle, the other operand (‘11111111’) is applied to the RWL lines of the same eight rows simultaneously. The combined duration for operand loading and RBL pre-charge is 63 ns.

Once the RWL signals are asserted, the MAC evaluation occurs within a 0.7 ns activation window. Extending this window further would fully discharge the RBL regardless of the MAC count, which is undesirable. The RBL node is modeled with a 200 fF load capacitance. As shown in the waveform, all comparator outputs (Vout0–Vout7) remain low for this case, corresponding to a decoded MAC value of “00000000”.

The complete operation, comprising operand loading, pre-charge, and evaluation—requires 63 ns in the proposed design. This results in an effective throughput of approximately 15.8 M operations/s, applicable to both MAC and all MAC-derived logic functions.

\subsection{Energy Calculations of MAC and Logic Operations}
The energy consumption for MAC operations is presented in Table~\ref{table:mac_energy}
, and the energy consumption for 1-bit logic operations AND, NOR, XOR, and addition is presented in Table~\ref{table:logic_energy}

\begin{table}[!h]
\centering
\caption{Energy Consumption for 8-Operand MAC}
\label{table:mac_energy}
\begin{tabular}{cc}
\toprule
MAC Count & RBL Energy (fJ) \\
\midrule
0 & 5.369 \\
1 & 119.3 \\
2 & 212.7 \\
3 & 288.5 \\
4 & 347.9 \\
5 & 391.6 \\
6 & 421.5 \\
7 & 440.7 \\
8 & 452.2 \\
\bottomrule
\end{tabular}
\end{table}

\begin{table}[!h]
\centering
\caption{Energy Consumption for 1-Bit Logic Operations}
\label{table:logic_energy}
\begin{tabular}{l c}
\toprule
Logic Operation & Energy Consumption (fJ) \\
\midrule
AND / Carry & 212.7 \\
NOR & 5.369 \\
XOR / Sum & 119.3 \\
\bottomrule
\end{tabular}
\end{table}

\subsection{Monte Carlo Analysis}
\begin{figure}[!t]
    \centering
    \includegraphics[width=\columnwidth]{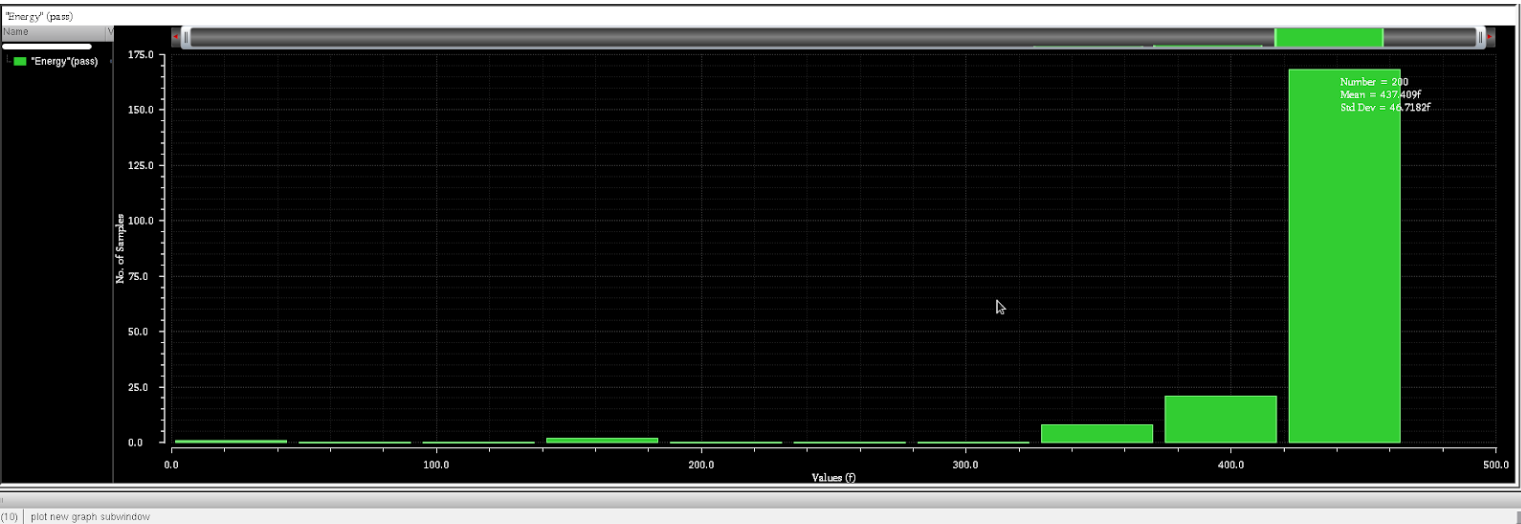} % change filename
    \caption{Monte Carlo histogram of energy consumption for MAC count = 8 over 200 samples.}
    \label{fig:monte_carlo}
\end{figure}

Monte Carlo analysis was performed for 200 samples for a MAC count of 8 and is presented in Fig.~\ref{fig:monte_carlo}. The mean energy consumption is found to be 437 fJ, and the standard deviation is 48.72 fJ.
Monte Carlo analysis in this work captures random device mismatch, which is the dominant source of variation during sensing. Although full Process Voltage Temperature (PVT) analysis is not included, the ordering of RBL voltage levels remains the same across process corners, meaning the decoder still sees a clear progression from MAC 0 to MAC 8. For larger PVT shifts, the reference thresholds of the decoder can be pre-characterized and adjusted based on corner simulations, ensuring correct operation without changing the overall architecture.

\begin{table}[H]
\centering
\caption{Comparison with Similar Work}
\label{table:comparison}

\footnotesize
\renewcommand{\arraystretch}{1.25}

\begin{tabular}{%
    >{\raggedright\arraybackslash}p{2.4cm} 
    >{\centering\arraybackslash}p{1.1cm} 
    >{\centering\arraybackslash}p{1.1cm} 
    >{\centering\arraybackslash}p{1.1cm} 
    >{\centering\arraybackslash}p{1.2cm}
}
\toprule
\textbf{Parameter} & \textbf{\cite{b2}} & \textbf{\cite{b9}} & \textbf{\cite{b10}} & \textbf{This work} \\
\midrule

\textbf{Node} &
28 nm &
28 nm &
45 nm &
90 nm \\

\textbf{Cell} &
6T &
8T &
8T, 8+T &
8T \\

\textbf{VDD} &
1 V &
0.65 V &
1.2 V &
1.8 V \\

\textbf{Number of operands} &
Multi (but limited) &
2 &
2 &
N \\

\textbf{Supported Boolean operations} &
AND, NOR &
AND, NAND, OR, NOR, XOR &
AND, NAND, OR, NOR, XOR &
MAC, AND, NAND, OR, NOR, XOR,\\ XNOR, 1-bit ADD. \\

\textbf{Frequency} &
370 MHz &
29 MHz (0.6 V),\\ 113 MHz (0.75 V) &
-- &
142.85 MHz \\

\textbf{Logic energy (fJ/bit)} &
-- &
3.7 (0.66 V) &
11.22--29.67 &
56.56 \\
\bottomrule
\end{tabular}
\end{table}

\subsection{Comparative Analysis}
Table V shows the comparison of this work with similar other work. This 8T SRAM-based IMC architecture distinguishes itself from existing designs through its unique ability to extract multiple logic operations from a single MAC computation. While most current SRAM IMC implementations focus exclusively on either MAC operations or individual logic functions, this paper simultaneously delivers AND, NOR, and XOR results by interpreting the MAC output through novel decoding techniques. This contrasts with traditional architectures that require separate dedicated circuits for each logic operation or multiple computation cycles to achieve similar functionality. 

 Since all Logic operations in the proposed design are inferred from a single MAC evaluation, no additional logic hardware is required. This reduces area and energy consumption which is important for edge-AI applications.

\section{Conclusion and future scope}
This paper presented an 8T SRAM-based in-memory computing architecture that unifies high-density storage with versatile arithmetic and logic processing. By decoupling the read and write paths, the proposed design overcomes the reliability challenges of conventional 6T cells while enabling parallel N-bit MAC operations and direct inference of fundamental logic functions from a single analog voltage measurement. Simulated in a 90 nm CMOS process, the 8×8 array achieves energy-efficient 8-bit MAC and logic operations at over 142 MHz and a latency of 0.7 ns and throughput of 15.8 M operations/s. The reported energy values focus on the MAC computation path, which consists mainly of IMC operation. Periphery overheads (write driver, decoder, precharge) follow predictable trends and will be analyzed in future extended work.

\end{document}